\documentclass[letterpaper,twocolumn,american,showpacs,pra,aps,superscriptaddress]{revtex4}
\usepackage[T1]{fontenc}
\usepackage[latin1]{inputenc}
\usepackage{bm}
\usepackage{amsmath}
\usepackage{babel}
\usepackage{graphicx}
\usepackage{amssymb}
\usepackage{hyperref}
\makeatletter
\usepackage{pifont}
\makeatother

\begin{document}

\title{Genuine quantum trajectories for non-Markovian processes}

\author{Heinz-Peter Breuer}

\email{breuer@physik.uni-freiburg.de}

\affiliation{Physikalisches Institut, Universit\"at Freiburg,
             Hermann-Herder-Str.~3, D-79104 Freiburg, Germany}

\date{\today}

\begin{abstract}
A large class of non-Markovian quantum processes in open systems
can be formulated through time-local master equations which are
not in Lindblad form. It is shown that such processes can be
embedded in a Markovian dynamics which involves a time dependent
Lindblad generator on an extended state space. If the state space
of the open system is given by some Hilbert space ${\mathcal{H}}$,
the extended state space is the triple Hilbert space
${\mathcal{H}}\otimes{\mathbb C}^3$ which is obtained by combining
the open system with a three state system. This embedding is used
to derive an unraveling for non-Markovian time evolution by means
of a stochastic process in the extended state space. The process
is defined through a stochastic Schr\"odinger equation which
generates genuine quantum trajectories for the state vector
conditioned on a continuous monitoring of an environment. The
construction leads to a continuous measurement interpretation for
non-Markovian dynamics within the framework of the theory of
quantum measurement.
\end{abstract}


\maketitle

\section{Introduction}
An open quantum system is a certain distinguished quantum system
which is coupled to another quantum system, its environment
\cite{TheWork}. A particularly simple way of describing an open
system is obtained in the Markovian approximation. In this
approximation all memory effects due to system-environment
correlations are neglected which usually leads to a Markovian
master equation, that is, to a linear first-order differential
equation for the reduced density matrix $\rho(t)$ of the open
system with a time independent generator. Generally, one demands
that the generator is in Lindblad form \cite{LINDBLAD}, which
follows from the requirements of the conservation of probability
and of the complete positivity of the dynamical map
\cite{GORINI1,GORINI2}.

A remarkable feature of Markovian master equations in Lindblad
form is given by the fact that they allow a stochastic
representation, also known as unraveling, by means of a stochastic
Schr\"odinger equation (SSE) for the state vector of the open
system \cite{CARMICHAEL,MOLMER,DUM,PLENIO,GISIN}. A SSE generates
the time evolution of the state vector which results from a
continuous monitoring of the environment of the system
\cite{WISEMAN1,WISEMAN2}. A specific realization
$\{|\psi(t)\rangle, t\geq 0\}$ of the SSE is called a quantum
trajectory: At each time $t \geq 0$ the open system is known to be
in a definite state $|\psi(t)\rangle$ under the condition that a
specific readout of the monitoring of the system's environment is
given. The reduced density matrix at time $t$ is therefore
obtained if one averages the quantity
$|\psi(t)\rangle\langle\psi(t)|$ over all possible quantum
trajectories. This means that the relation
$\rho(t)=E[|\psi(t)\rangle\langle\psi(t)|]$ holds, where the
symbol $E$ denotes the ensemble average or expectation value.

In the Markovian case it is thus true that the environment acts as
a quantum probe by which an indirect continuous observation of the
system is carried out. The description by means of a Markovian
master equation in Lindblad form is, however, only an
approximation which uses the assumption of short correlation
times. For strong couplings and low temperature environments
memory effects can lead to pronounced non-Markovian behavior.

It is sometimes argued that the treatment of non-Markovian
processes by means of master equations necessarily requires
solving integro-differential equations for the reduced density
matrix. Such equations arise, for example, in the application of
the Nakajima-Zwanzig projection operator technique
\cite{NAKAJIMA,ZWANZIG} which leads to dynamic equations involving
a retarded memory kernel and an integration over the past history
of the system.

However, the usage of another variant of the projection operator
method allows in many cases the derivation of approximate or even
exact non-Markovian master equations for the reduced density
matrix which are local in time. This method is known as the
time-convolutionless (TCL) projection operator technique
\cite{SHIBATA1,SHIBATA2,ROYER1,ROYER2}. The non-Markovian
character of the TCL master equation is reflected by the fact that
its generator may depend explicitly on time and may not be in
Lindblad form.

Time-local equations which are of the form of the TCL master
equation have also been derived by other means, e.~g., by path
integral and influence functional techniques
\cite{FEYNMAN,CALDEIRA}. A well known example is provided by the
exact equation of motion for a damped harmonic oscillator coupled
linearly to a bosonic reservoir \cite{HAAKE,HU,KARRLEIN}.

The fact that the TCL generator is generally not in Lindblad form
leads to several important mathematical and physical consequences.
In particular, a stochastic unraveling of the TCL master equation
of the form indicated above does not exist: Any such process will
automatically produce a master equation whose generator is in
Lindblad form. The question is therefore as to whether one can
develop a general method for the construction of stochastic
Schr\"odinger equations for non-Markovian dynamics which do have a
physical interpretation in terms of continuous measurements. It is
the purpose of this paper to show that this is indeed possible.

Our starting point is a time-local non-Markovian master equation
for the density matrix $\rho(t)$ on some Hilbert space
${\mathcal{H}}$ with a time dependent and bounded generator. It
will be demonstrated that the dynamics given by such a master
equation can always be embedded in a Markovian dynamics on an
appropriate extended state space. The non-Markovian dynamics thus
appears as part of a Markovian evolution in a larger state space.

If one chooses the extended state space as the Hilbert space of
the total system, consisting of open system plus environment, this
statement is of course trivial. However, it turns out that the
embedding can be realized in a fairly simple, much smaller state
space, namely in the tensor product space
${\mathcal{H}}\otimes{\mathbb C}^3$. In physical terms this is the
state space of a composite quantum system which results if one
combines the original open system on the state space
${\mathcal{H}}$ with a further auxiliary three state system
described by the state space ${\mathbb C}^3$. The open system
could be, for example, a damped quantum particle interacting with
a dissipative environment. The auxiliary system can then be
realized through an additional internal degree of freedom of the
particle which leads to a state space spanned by three basis
states.

It will be demonstrated that the dynamics in the extended state
space follows a Markovian master equation with a time dependent
generator in Lindblad form. The application of the standard
unraveling of Markovian master equations to the dynamics in the
extended state space therefore yields a stochastic unraveling for
the non-Markovian dynamics. The resulting SSE generates genuine
quantum trajectories which do admit a physical interpretation in
terms of a continuous observation carried out on an environment.
The construction thus gives rise to a consistent measurement
interpretation for non-Markovian evolution in full agreement with
the general setting of quantum measurement theory.

The paper is structured as follows. Section \ref{QMT-M} contains a
brief review of the continuous measurement theory for Markovian
dynamics. Time dependent generators in Lindblad form are
introduced in Sec.~\ref{LINDBLAD-GEN}, and Sec.~\ref{CMT-MARKOV}
treats the corresponding continuous measurement unraveling. The
quantum measurement theory for non-Markovian evolution is
developed in Sec.~\ref{QMT-NONMARKOV}. We introduce time-local
non-Markovian master equations in Sec.~\ref{TCL}. The embedding of
these equations in a Markovian dynamics is constructed in
Sec.~\ref{EMBEDDING}, whereas the derivation of the continuous
measurement unraveling is given in Sec.~\ref{SSE-NONMARKOV}. The
construction of the SSE and its physical interpretation are
illustrated by means of an example in Sec.~\ref{EXAMPLE}.

A series of interesting stochastic unravelings of non-Markovian
quantum dynamics is known in the literature. Section \ref{CONCLU}
contains a discussion of our results and of the relations to
alternative non-Markovian SSEs, as well as some conclusions.

\section{Quantum theory of Markovian dynamics}\label{QMT-M}

\subsection{Time dependent Lindblad generators}\label{LINDBLAD-GEN}
We consider a density matrix $W(t)$ on a state space
$\widetilde{\mathcal{H}}$ which obeys a master equation of the
form:
\begin{eqnarray} \label{LINDBLAD-EQ}
 \frac{d}{dt}W(t) &=& {\mathcal{L}}(t) W(t) \nonumber \\
 &\equiv& -i\left[ H(t),W(t) \right]
 + \sum_i J_i(t) W(t) J_i^{\dagger}(t) \nonumber \\
 &~& - \frac{1}{2} \sum_i \big\{ J_i^{\dagger}(t)J_i(t),W(t) \big\}.
\end{eqnarray}
The commutator with the Hamiltonian $H(t)$ represents the unitary
part of the evolution and the Lindblad operators $J_i(t)$ describe
the various decay channels of the system. In analogy to the
terminology used for classical master equations, the expressions
$J_iWJ_i^{\dagger}$ may be called gain terms, while the
expressions $\{J_i^{\dagger}J_i,W\}$, involving an
anti-commutator, may be referred to as loss terms.

Both the Hamiltonian $H(t)$ and the operators $J_i(t)$ are allowed
to depend on time $t$. The generator ${\mathcal{L}}(t)$ of the
master equation may thus be explicitly time dependent and does not
necessarily lead to a semigroup. We observe, however, that the
superoperator ${\mathcal{L}}(t_0)$ is in Lindblad form
\cite{LINDBLAD} for each fixed $t_0\geq 0$. This means that
${\cal{L}}(t_0)$ is in the form of the generator of a quantum
dynamical semigroup. The particular form of the generator derives
from the requirements of complete positivity and of the
conservation of the trace \cite{GORINI1,GORINI2}.

Under certain technical conditions which will be assumed to be
satisfied here, one concludes that Eq.~(\ref{LINDBLAD-EQ}) yields
a 2-parameter family of completely positive and trace preserving
maps $V(t,s)$ \cite{DAVIES,ALICKI}. These maps can be defined with
the help of the chronological time-ordering operator $T$ as
\begin{equation}
 V(t,s) = T \exp \left[ \int_s^t d\tau
 {\mathcal{L}}(\tau)\right], \qquad t \geq s \geq 0,
\end{equation}
and satisfy
\begin{equation}
 V(t,s) V(s,t') = V(t,t'), \qquad t \geq s \geq t'.
\end{equation}
In terms of these maps the solution of the master equation
(\ref{LINDBLAD-EQ}) at time $t$ can be written as
$W(t)=V(t,s)W(s)$, where $t \geq s \geq 0$. Thus, $V(t,s)$
propagates the density matrix at time $s$ to the density matrix at
time $t$.

Each $V(t,s)$ maps the space of density matrices into itself. This
means that $V(t,s)$ can be applied to any density matrix $W$ to
yield another density matrix $V(t,s)W$. The range of definition of
the maps $V(t,s)$ is thus the space of all density matrices and is
independent of time. Usually, one associates a Markovian master
equation with a time independent generator. We slightly generalize
this notion and refer to Eq.~(\ref{LINDBLAD-EQ}) as a Markovian
master equation with a time dependent Lindblad generator
${\mathcal{L}}(t)$.

\subsection{Stochastic unraveling and continuous measurement interpretation}
            \label{CMT-MARKOV}
As mentioned already in the Introduction the master equation
(\ref{LINDBLAD-EQ}) allows a stochastic unraveling through a
random process in the state space $\widetilde{\mathcal{H}}$. This
means that one can construct a stochastic dynamics for the state
vector $|\Phi(t)\rangle$ in $\widetilde{\mathcal{H}}$ which
reproduces the density matrix $W(t)$ with the help of the
expectation value
\begin{equation} \label{EXPEC}
 W(t) = E\big[|\Phi(t)\rangle\langle\Phi(t)|\big].
\end{equation}

To mathematically formulate this idea one writes a stochastic
Schr\"odinger equation (SSE) for the state vector
$|\Phi(t)\rangle$. An appropriate SSE for which the expectation
value (\ref{EXPEC}) leads to the master equation
(\ref{LINDBLAD-EQ}) is given by
\begin{eqnarray} \label{SSE}
 d|\Phi(t)\rangle &=& -i G(|\Phi(t)\rangle) dt \nonumber \\
 && + \sum_i \left[ \frac{J_i(t)|\Phi(t)\rangle}{||J_i(t)|\Phi(t)\rangle||}
 -|\Phi(t)\rangle \right] dN_i(t),
\end{eqnarray}
where we have introduced the nonlinear operator
\begin{equation} \label{DEF-G}
 G(|\Phi(t)\rangle) \equiv \left[ \hat{H}(t)+\frac{i}{2} \sum_i
 ||J_i(t)|\Phi(t)\rangle||^2 \right] |\Phi(t)\rangle.
\end{equation}
The term $-iG(|\Phi(t)\rangle)dt$ in Eq.~(\ref{SSE}), which is
proportional to the time increment $dt$, expresses the drift of
the process. This drift contribution obviously corresponds to the
nonlinear Schr\"odinger-type equation
\begin{eqnarray} \label{NSE}
 \frac{d}{dt} |\Phi(t)\rangle = -i G(|\Phi(t)\rangle),
\end{eqnarray}
whose linear part involves the non-hermitian Hamiltonian
\begin{equation} \label{H-HAT}
 \hat{H}(t) = H(t)-\frac{i}{2}\sum_i J_i^{\dagger}(t)J_i(t).
\end{equation}
The nonlinear part of the drift ensures that, although
$\hat{H}(t)$ is non-hermitian, the norm is conserved under the
deterministic time evolution given by Eq.~(\ref{NSE}).

The second term on the right-hand side of Eq.~(\ref{SSE})
represents a jump process leading to discontinuous changes of the
wave function, known as quantum jumps. These jumps are described
here with the help of the Poisson increments $dN_i(t)$ which
satisfy the relations:
\begin{eqnarray}
 dN_i(t)dN_j(t) &=& \delta_{ij}dN_i(t), \label{DEF-DN1} \\
 E[dN_i(t)] &=& ||J_i(t)|\Phi(t)\rangle||^2dt.
 \label{DEF-DN2}
\end{eqnarray}
According to Eq.~(\ref{DEF-DN1}) the increments $dN_i(t)$ are
random numbers which take the possible values $0$ or $1$.
Moreover, if $dN_i(t)=0$ for a particular $i$, we have $dN_j(t)=0$
for all $j \neq i$. The state vector then performs the jump
\begin{equation} \label{JUMP}
 |\Phi(t)\rangle \longrightarrow
 \frac{J_i(t)|\Phi(t)\rangle}{||J_i(t)|\Phi(t)\rangle||}.
\end{equation}
Thus we see that $N_i(t)$ is an integer-valued process which
counts the number of jumps of type $i$.

We infer from Eq.~(\ref{DEF-DN2}) that $dN_i(t)=1$ occurs with
probability $||J_i(t)|\Phi(t)\rangle||^2dt$. The jump described by
(\ref{JUMP}) thus takes place at a rate of
$||J_i(t)|\Phi(t)\rangle||^2$. The case $dN_i(t)=0$ for all $i$ is
realized with probability $1-\sum_i||J_i(t)|\Phi(t)\rangle||^2dt$.
In this case the state vector $|\Phi(t)\rangle$ follows the
deterministic drift described by Eq.~(\ref{NSE}).

Summarizing, the dynamics described by Eq.~(\ref{SSE}) yields a
piecewise deterministic process, i.~e., a random process whose
realizations consist of deterministic evolution periods
interrupted by discontinuous quantum jumps. Since both the
deterministic drift (\ref{NSE}) as well as the jumps (\ref{JUMP})
do not change the norm, the whole process preserves the norm of
the state vector.

The formulation of the dynamics by means of a SSE bears several
numerical advantages over the integration of the corresponding
density matrix equation (\ref{LINDBLAD-EQ}). This fact was the
original motivation for the development of stochastic wave
function methods in atomic physics and quantum optics (for an
example, see \cite{MOLMER2}). What is important in our context is
the fact that, additionally, the stochastic process given by the
SSE allows a physical interpretation in terms of a continuous
measurement which is carried out on an environment of the system.

To explain this point we consider a microscopic model in which the
open system is weakly coupled to a number of independent
reservoirs $R_i$, one reservoir for each value of the index $i$.
Each reservoir $R_i$ consists of bosonic modes $b_{i\lambda}$
which satisfy the commutation relations:
\begin{equation}
 \big[b_{i\lambda},b^{\dagger}_{j\mu}\big] =
 \delta_{ij}\delta_{\lambda\mu}.
\end{equation}
The Lindblad operators $J_i(t)$ appearing in the master equation
(\ref{LINDBLAD-EQ}) couple linearly to the reservoir operators
\begin{equation}
 B_i(t) = \sum_{\lambda} g_{i\lambda} e^{i(\omega_i-\omega_{i\lambda})t}
 b_{i\lambda},
\end{equation}
where the $\omega_i$ are certain system frequencies,
$\omega_{i\lambda}$ is the frequency of the mode $b_{i\lambda}$ of
reservoir $R_i$, and the $g_{i\lambda}$ are coupling constants.
Thus, the Hamiltonian of our model is taken to be:
\begin{equation}
 H_I(t) = H(t) + \frac{1}{\sqrt{\Gamma}}\sum_i \left[ J_i(t) B^{\dagger}_i(t)
 + J^{\dagger}_i(t) B_i(t) \right].
\end{equation}
We have included a factor $1/\sqrt{\Gamma}$, where $\Gamma$ is a
typical relaxation rate of the system which will be introduced
below. The combination $J_i/\sqrt{\Gamma}$ is therefore
dimensionless and the $B_i$ have the dimension of an inverse time,
choosing units such that $\hbar=1$.

The time evolution operator over the time interval $(t,t')$ will
be denoted by $U(t',t)$. The correlation function of reservoir
$R_i$ can be expressed through the spectral density $I(\omega)$,
which is assumed to be the same for all reservoirs:
\begin{equation}
 \langle 0| B_i(t')B_i^{\dagger}(t)|0\rangle =
 \int d\omega I(\omega) e^{i(\omega_i-\omega)(t'-t)}.
\end{equation}
Here, $|0\rangle$ denotes the vacuum state defined by
$b_{i\lambda}|0\rangle=0$ for all $i$ and all $\lambda$.

Suppose that the state of the combined system (open system plus
reservoirs $R_i$) at some time $t$ is given by
$|\Psi(t)\rangle=|\Phi(t)\rangle\otimes|0\rangle$. At time
$t'=t+\tau$ this state has evolved into the entangled state
$U(t',t)|\Psi(t)\rangle$. We consider $\tau$ to be a time
increment which is small compared to the time scale of the
systematic motion of the system, but large compared to the
correlation time of the reservoirs. Suppose further that at time
$t'$ a measurement of the quanta in the reservoir modes
$b_{i\lambda}$ is carried out. According to the standard theory of
quantum measurement \cite{BRAGINSKY} the detection of a quantum in
mode $b_{i\lambda}$ projects the state vector onto the state
$b^{\dagger}_{i\lambda}|0\rangle$. The open system's state
conditioned on this event thus becomes:
\begin{equation} \label{PROJ1}
 \frac{1}{\sqrt{p_{i\lambda}}}\langle 0 | b_{i\lambda}
 U(t',t)|\Psi(t)\rangle,
\end{equation}
where
\begin{equation}
 p_{i\lambda} = ||\langle 0 | b_{i\lambda}
 U(t',t)|\Psi(t)\rangle||^2
\end{equation}
is the corresponding probability. If, on the other hand, no
quantum is detected, one has to project the state vector onto the
vacuum state which yields the open system's conditioned state
\begin{equation} \label{PROJ0}
 \frac{1}{\sqrt{p_0}}\langle 0|U(t',t)|\Psi(t)\rangle,
\end{equation}
where
\begin{equation}
 p_0 = ||\langle 0|U(t',t)|\Psi(t)\rangle||^2
\end{equation}
is the probability that no quantum is detected.

In the Born-Markov approximation the above expression simplify
considerably. We take a constant spectral density
$I(\omega)=\Gamma/2\pi$, corresponding to the case of broad band
reservoirs with arbitrarily small correlation times. We further
neglect the Lamb shift contributions which lead to a
renormalization of the system Hamiltonian. The expression
(\ref{PROJ1}) then becomes (up to an irrelevant phase factor):
\begin{equation} \label{PROJ1BM}
 \frac{J_i(t)|\Phi(t)\rangle}{||J_i(t)|\Phi(t)\rangle||}.
\end{equation}
This expression is seen to be independent of $\lambda$. Therefore,
the total probability of observing a quantum in reservoir $R_i$
is:
\begin{equation}
 p_i = \sum_{\lambda} p_{i\lambda} = ||J_i(t)|\Phi(t)\rangle||^2
 \tau.
\end{equation}
The last two equation show that, conditioned on the detection of a
quantum in reservoir $R_i$, the system state carries out the jump
described in (\ref{JUMP}), and that these jumps occur at a rate
given by $||J_i(t)|\Phi(t)\rangle||^2$ [see Eq.~(\ref{DEF-DN2})].

For the case that no quantum is detected expression (\ref{PROJ0})
gives in the Born-Markov approximation:
\begin{equation} \label{PROJ0BM}
 |\Phi(t)\rangle - iG(|\Phi(t)\rangle)\tau,
\end{equation}
which leads to the drift contribution $-iG(|\Phi(t)\rangle)dt$ of
Eq.~(\ref{SSE}). The probability for this event is found to be
\begin{equation}
 p_0 = 1-\sum_{\lambda} p_{i\lambda}
 = 1-\sum_i||J_i(t)|\Phi(t)\rangle||^2 \tau.
\end{equation}

Considering that the detected quanta are annihilated on
measurement (quantum demolition measurement) we see that for both
alternatives described above the conditional state vector of the
combined system after time $\tau$ is again a tensor product of an
open system's state vector and the vacuum state of the
environment. Thus, we may repeat the measurement process after
each time increment $\tau$. In the limit of small $\tau$ we then
get a continuous measurement of the environment and a resulting
conditioned state vector of the open system that follows the SSE
(\ref{SSE}).

Summarizing, the SSE (\ref{SSE}) can be interpreted as resulting
from a continuous measurement of the quanta in the environment.
This measurement is an indirect measurement in which the jump
(\ref{JUMP}) of the state vector $|\Phi(t)\rangle$ describes the
measurement back action on the open system's state conditioned on
the detection of a quantum in reservoir $R_i$, while the nonlinear
Schr\"odinger equation (\ref{NSE}) yields the evolution of the
state vector under the condition that no quantum is detected. The
realizations of the process given by the SSE, i.e., the quantum
trajectories thus allow a clear physical interpretation in
accordance with the standard theory of quantum measurement.

\section{Quantum measurement interpretation of non-Markovian dynamics}
         \label{QMT-NONMARKOV}

\subsection{Time-local non-Markovian master equations} \label{TCL}
We investigate master equations for the density matrix $\rho(t)$
of an open system which are of the following general form:
\begin{eqnarray} \label{GEN-NON-MARKOV}
 \frac{d}{dt}\rho(t) &=& {\mathcal{K}}(t) \rho(t) \\
 &\equiv& -i\left[ H_S(t),\rho(t) \right] \nonumber \\
 &~& + \sum_{\alpha} \left[ C_{\alpha}(t) \rho(t) D_{\alpha}^{\dagger}(t)
 + D_{\alpha}(t) \rho(t) C_{\alpha}^{\dagger}(t) \right] \nonumber \\
 &~& -\frac{1}{2} \sum_{\alpha} \left\{ D_{\alpha}^{\dagger}(t) C_{\alpha}(t)
 + C_{\alpha}^{\dagger}(t) D_{\alpha}(t) , \rho(t) \right\}.
 \nonumber
\end{eqnarray}
The Hamiltonian $H_S(t)$, the $C_{\alpha}(t)$ and the
$D_{\alpha}(t)$ are given, possibly time dependent operators on
the state space ${\mathcal{H}}$ of the open system. The generator
${\mathcal{K}}(t)$ may thus again depend explicitly on time. The
master equation (\ref{GEN-NON-MARKOV}) is however local in time
since it does not contain a time integration over a memory kernel.
The structure of ${\mathcal{K}}(t)$ was taken to ensure that the
hermiticity and the trace of $\rho(t)$ are conserved. If we choose
$C_{\alpha}=D_{\alpha}\equiv J_i/\sqrt{2}$ the generator
${\mathcal{K}}(t)$ reduces to the form of a time dependent
Lindblad generator ${\mathcal{L}}(t)$. The Markovian master
equation (\ref{LINDBLAD-EQ}) is thus a special case of
Eq.~(\ref{GEN-NON-MARKOV}) which will be referred to as
non-Markovian master equation.

With an appropriate choice for the Hamiltonian $H_S(t)$ and the
operators $C_{\alpha}(t)$ and $D_{\alpha}(t)$, a large variety of
physical phenomena can be described by master equations of the
form (\ref{GEN-NON-MARKOV}). For example, as mentioned in the
Introduction a master equation of this form arises when applying
the TCL projection operator technique \cite{SHIBATA1,SHIBATA2} to
the dynamics of open system. The basic idea underlying this
technique is to remove the memory kernel from the equations of
motion by the introduction of the backward propagator. Under the
condition of factorizing initial conditions one then finds a
homogeneous master equation with a time-local generator
${\mathcal{K}}(t)$. The latter can be determined explicitly
through a systematic perturbation expansion in terms of ordered
cumulants \cite{ROYER1,ROYER2}. Specific examples are the TCL
master equations describing spin relaxation
\cite{CHANG,DESPOSITO}, the spin-boson model \cite{ANNPHYS},
systems coupled to a spin bath \cite{PDP-EPJD,BURGARTH}, charged
particles interacting with the electromagnetic field \cite{BREMS},
and the atom laser \cite{FALLER}.

Moreover, several exact time-local master equations of the form
(\ref{GEN-NON-MARKOV}) are known in the literature which have been
derived by other means. Examples are provided by the master
equations for non-Markovian quantum Brownian motion
\cite{HAAKE,HU,KARRLEIN} and for the nonperturbative decay of
atomic systems, which will be discussed in Sec.~\ref{EXAMPLE}.

The existence of a homogeneous, time-local master equation
requires, in general, that the initial state of the total system
represents a tensor product state. For simplicity we restrict
ourselves to this case since we intend to develop stochastic
unravelings for pure states of the reduced open system.

Due to its explicit time dependence the generator
${\mathcal{K}}(t)$ of the master equation (\ref{GEN-NON-MARKOV})
does of course not lead to a semigroup. But even for a fixed $t
\geq 0$ the superoperator ${\mathcal{K}}(t)$ is, in general, not
in Lindblad form, by contrast to the property of the generator
${\mathcal{L}}(t)$ of the master equation (\ref{LINDBLAD-EQ}). To
make this point more explicit we introduce operators
$E_{\alpha}(t)=C_{\alpha}(t)-D_{\alpha}(t)$ and rewrite
Eq.~(\ref{GEN-NON-MARKOV}) as:
\begin{equation} \label{GEN-NON-MARKOV2}
 \frac{d}{dt}\rho(t) = -i\left[ H_S(t),\rho(t) \right]
 + {\mathcal{D}}_1(t)\rho(t) + {\mathcal{D}}_2(t)\rho(t),
\end{equation}
where we have defined the superoperators
\begin{eqnarray*}
 {\mathcal{D}}_1(t)\rho &\equiv& \sum_{\alpha} \left[
 C_{\alpha}(t) \rho C_{\alpha}^{\dagger}(t)
 -\frac{1}{2}\left\{C_{\alpha}^{\dagger}(t)C_{\alpha}(t),\rho\right\}
 \right. \nonumber \\
 &~& \qquad \left. + D_{\alpha}(t) \rho D_{\alpha}^{\dagger}(t)
 -\frac{1}{2}\left\{D_{\alpha}^{\dagger}(t)D_{\alpha}(t),\rho\right\}
 \right], \\
 {\mathcal{D}}_2(t)\rho &\equiv& -\sum_{\alpha}
 \left[ E_{\alpha}(t) \rho E_{\alpha}^{\dagger}(t)
 - \frac{1}{2}\left\{E_{\alpha}^{\dagger}(t)E_{\alpha}(t),\rho\right\}
 \right].
\end{eqnarray*}
One observes that ${\mathcal{D}}_1(t)$ is in Lindblad form, while
${\mathcal{D}}_2(t)$ is not: The superoperator
${\mathcal{D}}_2(t)$ carries an overall minus sign and violates
therefore the complete positivity of the generator.

Of course, we will assume in the following that the dynamics given
by Eq.~(\ref{GEN-NON-MARKOV}) yields a dynamical map $\rho(0)
\mapsto \rho(t)$ for all times considered, that is, we suppose
that Eq.~(\ref{GEN-NON-MARKOV}) describes the evolution of true
density matrices at time $t=0$ into true density matrices at time
$t>0$. However, this assumption does not imply that the
propagation of an arbitrary positive matrix at time $t$ with the
help of the master equation (\ref{GEN-NON-MARKOV}) necessarily
leads to a positive matrix for future times. This is only
guaranteed if we propagate a density matrix $\rho(t)$ which
results from the time evolution over the previous interval
$(0,t)$.

A further important consequence of the form of the master equation
(\ref{GEN-NON-MARKOV}) is that it does not allow a stochastic
unraveling of the type developed in Sec.~\ref{CMT-MARKOV}. Any
unraveling of this kind would automatically produce a master
equation with a time dependent Lindblad generator. Trying to
construct an unraveling which leads to the contribution
${\mathcal{D}}_2(t)\rho(t)$ of the master equation, one would find
a process with negative transitions rates, which is both
unphysical and mathematically inconsistent.

\subsection{Markovian embedding of non-Markovian dynamics}\label{EMBEDDING}
To overcome the difficulties in the development of a stochastic
representation we are going to employ an interesting general
feature of the non-Markovian master equation
(\ref{GEN-NON-MARKOV}): Even if the generator ${\mathcal{K}}(t)$
is not in Lindblad form, it is always possible to construct an
embedding of the non-Markovian dynamics in a Markovian evolution
on a suitable extended state space. The precise formulation of
this statement and its proof will be given in the following.

The extended state space is obtained by combining the original
open system on the state space ${\mathcal{H}}$ with another
auxiliary quantum system. The auxiliary system is a three state
system whose state space ${\mathbb C}^3$ is spanned by three basis
states $|1\rangle$, $|2\rangle$ and $|3\rangle$:
\begin{equation}
 {\mathbb C}^3 = {\mathrm{span}}\{ |1\rangle, |2\rangle,
 |3\rangle\}.
\end{equation}
The Hilbert space $\widetilde{\mathcal{H}}$ of the combined system
then becomes the triple Hilbert space
\begin{equation}
 \widetilde{\mathcal{H}} = {\mathcal{H}} \otimes {\mathbb C}^3
 \cong {\mathcal{H}}_1\oplus{\mathcal{H}}_2\oplus{\mathcal{H}}_3.
\end{equation}
The extended state space is thus given by the tensor product of
${\mathcal{H}}$ and ${\mathbb C}^3$, which in turn is isomorphic
to the orthogonal sum of three copies ${\mathcal{H}}_1$,
${\mathcal{H}}_2$, ${\mathcal{H}}_3$ of ${\mathcal{H}}$. This
means that states $|\Phi\rangle$ in $\widetilde{\mathcal{H}}$ take
the general form:
\begin{equation} \label{REPR}
 |\Phi\rangle = |\psi_1\rangle \otimes |1\rangle
 + |\psi_2\rangle \otimes |2\rangle + |\psi_3\rangle \otimes
 |3\rangle,
\end{equation}
where $|\psi_k\rangle \in {\mathcal{H}}$ for $k=1,2,3$. As a
possible physical realization of the extended state space on may
think of ${\mathcal{H}}$ as the state space of a damped quantum
particle with an additional internal degree of freedom which can
be represented by a three level system.

We now regard Eq.~(\ref{LINDBLAD-EQ}) as an equation of motion on
the triple Hilbert space, that is, $W(t)$ is considered as a
density matrix on $\widetilde{\mathcal{H}}$ governed by a master
equation with time dependent Lindblad generator
${\mathcal{L}}(t)$. On the other hand, $\rho(t)$ is a density
matrix on ${\mathcal{H}}$ satisfying the given non-Markovian
master equation (\ref{GEN-NON-MARKOV}). We will assume in the
following that the operators $C_{\alpha}(t)$ and $D_{\alpha}(t)$
are bounded.

\begin{figure}[htb]
\includegraphics[width=0.9\linewidth]{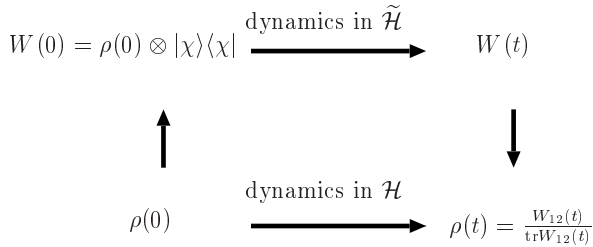}
\caption{Illustration of the embedding theorem: The initial
density matrix $\rho(0)$ evolves into $\rho(t)$ according to the
given non-Markovian master equation (\ref{GEN-NON-MARKOV}). This
evolution can be embedded in a Markovian dynamics on the extended
state space $\widetilde{\mathcal{H}}={\mathcal{H}}\otimes{\mathbb
C}^3$ in which the density matrix $W(0)$ evolves into $W(t)$
following the master equation (\ref{LINDBLAD-EQ}).\label{fig1}}
\end{figure}

Our aim is to show that by an appropriate choice of the
Hamiltonian $H(t)$ and of the Lindblad operators $J_{i}(t)$ in
Eq.~(\ref{LINDBLAD-EQ}) one can always achieve that the density
matrix $\rho(t)$ on ${\mathcal{H}}$ is connected to the density
matrix $W(t)$ on $\widetilde{\mathcal{H}}$ by means of the
relation (${\mathrm{tr}}$ denotes the trace):
\begin{equation} \label{DEF-RHO}
 \rho(t) = \frac{W_{12}(t)}{{\mathrm{tr}}W_{12}(t)}.
\end{equation}
Here, we have defined
\begin{equation}
 W_{12}(t) = \langle 1 | W(t) | 2 \rangle,
\end{equation}
which is an operator acting on ${\mathcal{H}}$. We can regard this
operator $W_{12}(t)$ as a matrix which is formed by the coherences
(off-diagonal elements) of $W(t)$ between states from the subspace
${\mathcal{H}}_1$ and states from the subspace ${\mathcal{H}}_2$.

This is the embedding theorem. It states that the non-Markovian
dynamics of $\rho(t)$ can be expressed through the time evolution
of a certain set of coherences $W_{12}(t)$ of a density matrix
$W(t)$ on the extended space which follows a Markovian dynamics
(see Fig.~\ref{fig1}).

To proof the embedding theorem we first demonstrate that the
relation (\ref{DEF-RHO}) can be achieved to hold at time $t=0$. If
$\rho(0)$ is any initial density matrix on ${\mathcal{H}}$ we
define a corresponding density matrix on $\widetilde{\mathcal{H}}$
by
\begin{eqnarray} \label{DEF-W0}
 W(0) &\equiv& \rho(0) \otimes \frac{1}{2} \big[ |1\rangle\langle 1|
 + |2\rangle\langle 2| + |1\rangle\langle 2| + |2\rangle\langle 1|
 \big] \nonumber \\
 &=& \rho(0) \otimes |\chi\rangle\langle\chi|,
\end{eqnarray}
where
\begin{eqnarray} \label{CHI}
 |\chi\rangle = \frac{1}{\sqrt{2}} \big[ |1\rangle + |2\rangle \big]
\end{eqnarray}
is a state vector of the auxiliary system. We thus obtain $W(0)$
by combining the open system in the state $\rho(0)$ with the three
state system in the pure state $|\chi\rangle\langle\chi|$. It is
obvious that $W(0)$ given by Eq.~(\ref{DEF-W0}) is a true density
matrix on the triple Hilbert space $\widetilde{\mathcal{H}}$,
i.~e., we have $W(0) \geq 0$ and ${\mathrm{tr}}W(t)=1$. Moreover,
Eq.~(\ref{DEF-W0}) yields:
\begin{equation}
 \frac{W_{12}(0)}{{\mathrm{tr}}W_{12}(0)}
 = \frac{\langle 1 |W(0)|2\rangle}{{\mathrm{tr}}\langle 1|W(0)|2\rangle}
 = \frac{\frac{1}{2}\rho(0)}{\frac{1}{2}{\mathrm{tr}}\rho(0)}
 = \rho(0),
\end{equation}
which is Eq.~(\ref{DEF-RHO}) at time $t=0$.

To show that the relation (\ref{DEF-RHO}) is valid for all times
$t \geq 0$ we have to demonstrate that the right-hand side of this
relation satisfies the non-Markovian master equation
(\ref{GEN-NON-MARKOV}), provided the Lindblad operators $J_i(t)$
in Eq.~(\ref{LINDBLAD-EQ}) are chosen appropriately. To simplify
the presentation we first treat the case that the master equation
(\ref{GEN-NON-MARKOV}) only involves a single operator $C(t)$ and
$D(t)$. Thus we write, suppressing the time arguments,
\begin{eqnarray} \label{NMSA}
 \frac{d}{dt} \rho
 &=& -i [H_S,\rho] + C \rho D^{\dagger} + D \rho C^{\dagger}
 \nonumber \\
 &~& -\frac{1}{2} \left\{
 D^{\dagger}C + C^{\dagger}D , \rho \right\}.
\end{eqnarray}

\begin{figure}[htb]
\includegraphics[width=0.9\linewidth]{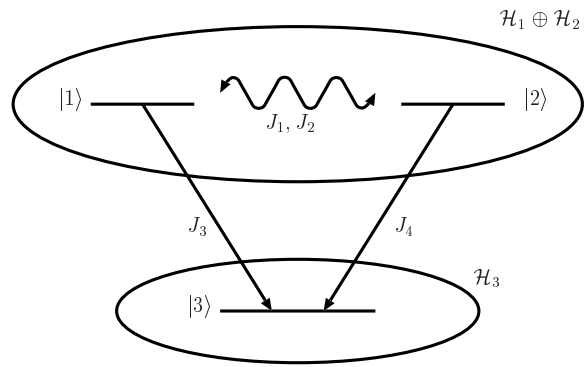}
\caption{The extended state space ${\mathcal{H}}\otimes{\mathbb
C}^3={\mathcal{H}}_1\oplus{\mathcal{H}}_2\oplus{\mathcal{H}}_3$
and the action of the operators $J_i$ defined in
Eqs.~(\ref{DEF-J1})-(\ref{DEF-J4}).\label{fig2}}
\end{figure}

We define four time dependent Lindblad operators for the master
equation (\ref{LINDBLAD-EQ}):
\begin{eqnarray}
 J_1(t) &=& C(t) \otimes |1\rangle\langle 1| + D(t) \otimes |2\rangle\langle 2|,
 \label{DEF-J1} \\
 J_2(t) &=& D(t) \otimes |1\rangle\langle 1| + C(t) \otimes |2\rangle\langle 2|,
 \label{DEF-J2} \\
 J_3(t) &=& \Omega(t) \otimes |3\rangle\langle 1|,
 \label{DEF-J3} \\
 J_4(t) &=& \Omega(t) \otimes |3\rangle\langle 2|,
 \label{DEF-J4}
\end{eqnarray}
and the Hamiltonian:
\begin{eqnarray} \label{DEF-H}
 H(t) &=& H_S(t) \otimes \big[ |1\rangle\langle 1| + |2\rangle\langle 2|
 + |3\rangle\langle 3| \big] \nonumber \\
 &\equiv& H_S(t) \otimes I_3,
\end{eqnarray}
where $I_3$ denotes the unit operator on the auxiliary space
${\mathbb C}^3$. In Eqs.~(\ref{DEF-J3}) and (\ref{DEF-J4}) we have
introduced a time dependent operator $\Omega(t)$ on
${\mathcal{H}}$ which will be defined below.

According to the definitions (\ref{DEF-J1}) and (\ref{DEF-J2}) the
operators $J_1$ and $J_2$ leave invariant the subspace
${\mathcal{H}}_1\oplus{\mathcal{H}}_2$ which contains the states
of the form
$|\psi_1\rangle\otimes|1\rangle+|\psi_2\rangle\otimes|2\rangle$.
The operators $J_3$ and $J_4$ defined in Eqs.~(\ref{DEF-J3}) and
(\ref{DEF-J4}) induce transitions $|1\rangle\rightarrow|3\rangle$
and $|2\rangle\rightarrow|3\rangle$ between the states of the
auxiliary three state system. The extended state space and the
action of the Lindblad operators are illustrated in
Fig.~\ref{fig2}.

We have from the above definitions (suppressing again the time
arguments):
\begin{eqnarray}
 J_1^{\dagger}J_1 &=& C^{\dagger}C \otimes |1\rangle\langle 1|
 + D^{\dagger}D \otimes |2\rangle\langle 2|, \\
 J_2^{\dagger}J_2 &=& D^{\dagger}D \otimes |1\rangle\langle 1|
 + C^{\dagger}C \otimes |2\rangle\langle 2|, \\
 J_3^{\dagger}J_3 &=& \Omega^{\dagger}\Omega
 \otimes |1\rangle\langle 1|, \\
 J_4^{\dagger}J_4 &=& \Omega^{\dagger}\Omega
 \otimes |2\rangle\langle 2|,
\end{eqnarray}
and:
\begin{equation} \label{SUMI-JI}
 \sum_{i=1}^4  J_i^{\dagger}J_i = \left(\Omega^{\dagger}\Omega
 + C^{\dagger}C + D^{\dagger}D \right) \otimes \left(
 |1\rangle\langle 1| + |2\rangle\langle 2| \right).
\end{equation}
With the Lindblad operators defined in
Eqs.~(\ref{DEF-J1})-(\ref{DEF-J4}) and with the Hamiltonian given
by Eq.~(\ref{DEF-H}) the master equation (\ref{LINDBLAD-EQ}) leads
to:
\begin{eqnarray} \label{EQ-W-1}
 \frac{d}{dt} W_{12}
 &=& -i \langle 1 | [H,W] | 2 \rangle
 + \sum_{i=1}^4 \langle 1 | J_i W J_i^{\dagger} | 2 \rangle
 \nonumber \\
 && -\frac{1}{2} \sum_{i=1}^4
 \langle 1 | \big\{ J_i^{\dagger}J_i,W \big\} | 2 \rangle.
\end{eqnarray}
On using Eq.~(\ref{SUMI-JI}) we find
\begin{equation} \label{LOSSTERM}
 \sum_{i=1}^4 \langle 1 | \big\{ J_i^{\dagger}J_i,W \big\} | 2 \rangle
 = \left\{ \Omega^{\dagger}\Omega
 + C^{\dagger}C + D^{\dagger}D , W_{12} \right\},
\end{equation}
while Eqs.~(\ref{DEF-J1})-(\ref{DEF-H}) yield
\begin{eqnarray}
 \langle 1 | J_1 W J_1^{\dagger} | 2 \rangle
 &=& C W_{12} D^{\dagger}, \label{REL1} \\
 \langle 1 | J_2 W J_2^{\dagger} | 2 \rangle
 &=& D W_{12} C^{\dagger}, \label{REL2} \\
 \langle 1 | J_3 W J_3^{\dagger} | 2 \rangle
 &=& \langle 1 | J_4 W J_4^{\dagger} | 2 \rangle = 0,
 \label{REL34} \\
 \langle 1 | [H,W] | 2 \rangle &=& [H_S,W_{12}]. \label{REL5}
\end{eqnarray}

Employing the relations (\ref{LOSSTERM})-(\ref{REL5}) in
Eq.~(\ref{EQ-W-1}) we arrive at:
\begin{eqnarray} \label{EQ-W-2}
 \frac{d}{dt} W_{12}
 &=& -i [H_S,W_{12}] + C W_{12} D^{\dagger} + D W_{12} C^{\dagger}
 \nonumber \\
 && -\frac{1}{2} \left\{ \Omega^{\dagger}\Omega
 + C^{\dagger}C + D^{\dagger}D , W_{12} \right\}.
\end{eqnarray}
Here we see the reason for our choice of the extended state space
and of the Lindblad operators $J_i$. The operators $J_1$ and $J_2$
have been chosen in such a way that $C$ acts from the left and
$D^{\dagger}$ from the right on the coherences $W_{12}$, or vice
versa [see Eqs.~(\ref{REL1}) and (\ref{REL2})]. On the other hand,
since $J_3$ and $J_4$ induce transitions into the state
$|3\rangle$ of the auxiliary system, the corresponding gain terms
$J_3WJ^{\dagger}_3$ and $J_4WJ^{\dagger}_4$ of the master equation
(\ref{LINDBLAD-EQ}) do not contribute towards the equation of
motion (\ref{EQ-W-2}) for the coherences $W_{12}$ [see
Eq.~(\ref{REL34})]. The subspace ${\mathcal{H}}_3$ of the extended
state space plays the role of a sink which will be used now to
achieve that the loss terms of the master equation come out
correctly.

We observe that Eq.~(\ref{EQ-W-2}) is already of a form which is
similar to the desired master equation (\ref{NMSA}). These
equations differ, however, with respect to the structure of the
loss terms which are given by the terms containing the
anti-commutator. To get an equation of motion of the desired form
we now choose $\Omega$ to be a solution of the equation
\begin{eqnarray} \label{OMEGA-EQ}
 \Omega^{\dagger}\Omega + C^{\dagger}C + D^{\dagger}D
 = a I + D^{\dagger}C + C^{\dagger}D,
\end{eqnarray}
which is equivalent to
\begin{eqnarray} \label{OMEGA-DEF}
 \Omega^{\dagger}\Omega
 = a I - (C-D)^{\dagger}(C-D).
\end{eqnarray}
Here, $I$ denotes the unit operator on ${\mathcal{H}}$ and
$a=a(t)$ is a time dependent non-negative number. Since
$\Omega^{\dagger}\Omega$ is a positive operator a solution
$\Omega$ of Eq.~(\ref{OMEGA-DEF}) exists under the condition that
the right-hand side of Eq.~(\ref{OMEGA-DEF}) is also a positive
operator. Thus, $a$ must be chosen in such a way that
\begin{equation} \label{INEQ}
 a \geq ||(C-D)|\psi\rangle||^2
\end{equation}
for all normalized state vectors $|\psi\rangle$ in
${\mathcal{H}}$. We note that it is at this point that the
assumption of bounded operators enters our construction. To make a
definite choice we define $a$ to be the largest eigenvalue of the
positive operator $(C-D)^{\dagger}(C-D)$. This definition ensures
that the inequality (\ref{INEQ}) is satisfied and that a solution
$\Omega$ of Eq.~(\ref{OMEGA-DEF}) exists.

The solution of Eq.~(\ref{OMEGA-DEF}) is, in general, not unique.
If $\Omega$ is a solution, then also $U\Omega$, where $U$ is an
arbitrary unitary operator. Changing $\Omega$ into $U\Omega$ does,
however, not influence the equation of motion (\ref{EQ-W-2}) since
only the combination $\Omega^{\dagger}\Omega$ enters this
equation. In the language of quantum measurement theory the
transformation $\rho \mapsto \Omega\rho\Omega^{\dagger}$ is called
a quantum operation \cite{KRAUS}. It describes the change of a
density matrix $\rho$ under a generalized measurement whose
outcome occurs with probability
${\mathrm{tr}}(\Omega^{\dagger}\Omega\rho)$. We can thus say that
the unitary operator $U$, expressing our freedom in the choice of
$\Omega$, affects the change of the system state $\rho$, but not
the probability of its occurrence.

Substituting Eq.~(\ref{OMEGA-EQ}) into Eq.~(\ref{EQ-W-2}) we get:
\begin{eqnarray} \label{EQ-W-3}
 \frac{d}{dt} W_{12}
 &=& -i [H_S,W_{12}] + C W_{12} D^{\dagger} + D W_{12} C^{\dagger}
 \nonumber \\
 && -\frac{1}{2} \left\{
 D^{\dagger}C + C^{\dagger}D , W_{12} \right\} - a W_{12}.
\end{eqnarray}
We conclude from this equation that the trace of $W_{12}$
satisfies the equation
\begin{equation} \label{TRACE}
 \frac{d}{dt} {\mathrm{tr}} W_{12}=-a{\mathrm{tr}}W_{12}.
\end{equation}
Using this fact as well as Eq.~(\ref{EQ-W-3}) one immediately
demonstrates that the expression on the right-hand side of
Eq.~(\ref{DEF-RHO}) satisfies the desired master equation
(\ref{NMSA}), which concludes the proof of the embedding theorem.

The general case of an arbitrary number of operators
$C_{\alpha}(t)$ and $D_{\alpha}(t)$ in Eq.~(\ref{GEN-NON-MARKOV})
can be treated in a similar way. To this end, one has to
re-introduce the index $\alpha$ and to carry out a summation over
$\alpha$ in the equations of motion. Thus, for each value of
$\alpha$ we have a corresponding $\Omega_{\alpha}(t)$ and an
$a_{\alpha}(t)$, as well as four $J_{i\alpha}(t)$, $i=1,2,3,4$.

Finally we note that according to Eq.~(\ref{EQ-W-3}) the operator
$W_{21}\equiv\langle 2|W|1\rangle=W_{12}^{\dagger}$ satisfies the
same differential equation as $W_{12}$. Since also
$W_{21}(0)=W_{12}(0)$, we conclude that $W_{21}(t)=W_{12}(t)$ for
all times. It follows that we can write for any operator $A$ on
${\mathcal{H}}$:
\begin{equation}
 {\mathrm{tr}} \left\{ A \rho(t) \right\}
 = \frac{{\mathrm{tr}}\{(A \otimes \sigma_x)W(t)\}}
 {{\mathrm{tr}}\{(I\otimes \sigma_x)W(t)\}},
\end{equation}
where
\begin{equation}
 \sigma_x = |1\rangle\langle 2| + |2\rangle\langle 1|
\end{equation}
is an operator on the auxiliary state space. This shows that the
expectation value of all observables $A$ in the state $\rho(t)$
can be determined through measurements on the state $W(t)$ of the
extended system.

\subsection{Stochastic unraveling for non-Markovian processes}
            \label{SSE-NONMARKOV}
The embedding of the previous section enables us to construct a
stochastic unraveling for the non-Markovian dynamics given by the
master equation (\ref{GEN-NON-MARKOV}). Since the master equation
governing $W(t)$ involves a time dependent Lindblad generator we
can use the SSE developed in Sec.~\ref{CMT-MARKOV} for this
purpose.

The SSE (\ref{SSE}) generates a stochastic process for the state
vector $|\Phi(t)\rangle$ in the triple Hilbert space. Employing
the representation (\ref{REPR}) we write
\begin{equation} \label{REPR-T}
 |\Phi(t)\rangle = |\psi_1(t)\rangle \otimes |1\rangle
 + |\psi_2(t)\rangle \otimes |2\rangle + |\psi_3(t)\rangle \otimes
 |3\rangle.
\end{equation}
As shown in Sec.~\ref{CMT-MARKOV} the density matrix $W(t)$ on the
extended state space is reproduced through the expectation value
$W(t)=E[|\Phi(t)\rangle\langle\Phi(t)|]$, and the norm of the
state vector is exactly conserved during the stochastic evolution:
\begin{eqnarray} \label{NORM}
 \lefteqn{ \langle\Phi(t)|\Phi(t)\rangle } \nonumber \\
 &\equiv& \langle\psi_1(t)|\psi_1(t)\rangle
 + \langle\psi_2(t)|\psi_2(t)\rangle
 + \langle\psi_3(t)|\psi_3(t)\rangle \nonumber \\
 &=& 1.
\end{eqnarray}

In accordance with Eq.~(\ref{DEF-W0}) the initial state of the
process is taken to be of the form:
\begin{equation}
 |\Phi(0)\rangle = |\varphi\rangle \otimes |\chi\rangle
 = \frac{1}{\sqrt{2}} \left[ |\varphi\rangle\otimes|1\rangle
 + |\varphi\rangle\otimes|2\rangle \right],
\end{equation}
where $|\varphi\rangle$ is a normalized random state vector in
${\mathcal{H}}$, $\langle\varphi|\varphi\rangle=1$, and
$|\chi\rangle$ is the fixed state vector of the auxiliary three
state system defined in (\ref{CHI}). We thus have
$|\psi_1(0)\rangle=|\psi_2(0)\rangle=\frac{1}{\sqrt{2}}|\varphi\rangle$
and $|\psi_3(0)\rangle=0$. Hence, Eq.~(\ref{DEF-RHO}) at time
$t=0$ gives
\begin{equation}
 \rho(0) = E[|\varphi\rangle\langle\varphi|].
\end{equation}

The embedding theorem now reveals that $\rho(t)$ is obtained from
the stochastic evolution with the help of the relation [see
Eq.~(\ref{DEF-RHO})]:
\begin{equation} \label{RHO-STOCH}
 \rho(t)
 = \frac{E(|\psi_1(t)\rangle\langle\psi_2(t)|)}
 {E(\langle\psi_2(t)|\psi_1(t)\rangle)}.
\end{equation}
Thus we have constructed a stochastic unraveling for the
non-Markovian dynamics. It is important to realize that our
construction leads to an unraveling through a normalized
stochastic state vector $|\Phi(t)\rangle$ and that the process
allows a definite physical interpretation in terms of a continuous
measurement, as has been discussed in Sec.~\ref{CMT-MARKOV}.

We remark that for the case $C=D$ the given master equation
(\ref{NMSA}) is already in time dependent Lindblad form. Our
construction then yields $a(t)=0$ [see inequality (\ref{INEQ})]
and $J_1=J_2$, as well as $J_3=J_4=0$. This means that the jump
operators $J_1$ and $J_2$ are identical and that the decay
channels $J_3$ and $J_4$ are closed. It follows that
$|\psi_3(t)\rangle\equiv 0$ and that $|\psi_1\rangle$ evolves in
exactly the same way as $|\psi_2\rangle$, that is, we have
$|\psi_1(t)\rangle \equiv |\psi_2(t)\rangle$. Equation
(\ref{RHO-STOCH}) thus becomes
$\rho(t)=2E(|\psi_1(t)\rangle\langle\psi_1(t)|)$. Note that the
factor $2$ is due to the normalization condition (\ref{NORM})
which yields
$\langle\psi_1|\psi_1\rangle=\langle\psi_2|\psi_2\rangle=1/2$. In
the case $C=D$ our construction therefore reduces automatically to
the standard unraveling of a master equation in Lindblad form.

\section{Example}\label{EXAMPLE}
As an example we discuss a model for the non-Markovian decay of a
two state system into the vacuum of a bosonic bath. The model
serves to illustrate how to construct the embedding in a Markovian
dynamics, and how to interpret physically the quantum trajectories
generated by the resulting SSE.

\subsection{Construction of the process}\label{PROC-JC}
The interaction picture master equation of the model is given by
\begin{eqnarray} \label{JC-MODEL}
 \frac{d}{dt}\rho(t) &=& -i [H_S(t),\rho(t)] \\
 &~& +\gamma(t)\left( \sigma_-\rho(t)\sigma_+
 -\frac{1}{2}\left\{ \sigma_+\sigma_-,\rho(t) \right\} \right),
 \nonumber
\end{eqnarray}
where
\begin{equation}
 H_S(t) = \frac{1}{2}S(t)\sigma_+\sigma_-.
\end{equation}
This is an exact master equation for the nonperturbative decay of
a two state system with excited state $|e\rangle$ and ground state
$|g\rangle$, which interacts with a bosonic bath \cite{TheWork}.
$\sigma_+=|e\rangle\langle g|$ and $\sigma_-=|g\rangle\langle e|$
are the usual raising and lowering operators of the two state
system. These operators couple linearly to the bath through an
interaction Hamiltonian of the form
$\sigma_-Q^{\dagger}(t)+\sigma_+Q(t)$, where $Q(t)$ is a bath
operator depending linearly on the annihilation operators of the
bath modes.

The real functions $S(t)$ and $\gamma(t)$ are determined by the
vacuum correlation function $\langle
0|Q(t)Q^{\dagger}(t_1)|0\rangle$ of the bath. An example will be
discussed below [see Eq.~(\ref{CORR-JC})]. The function $S(t)$
describes a time dependent renormalization of the system
Hamiltonian induced by the coupling to the bath (Lamb shift).
Under the condition $\gamma(t)\geq0$ one can interpret the
function $\gamma(t)$ as a time dependent decay rate of the excited
state. But for certain spectral densities $\gamma(t)$ may become
negative in certain time intervals such that the master equation
(\ref{JC-MODEL}) is not in Lindblad form and the generator is not
completely positive.

The master equation (\ref{JC-MODEL}) can however always be brought
into the form of the non-Markovian master equation (\ref{NMSA}) by
means of the definitions:
\begin{eqnarray}
 C(t) &=& \sqrt{\frac{|\gamma(t)|}{2}} \sigma_-, \\
 D(t) &=& \sqrt{\frac{|\gamma(t)|}{2}} {\mathrm{sign}} \gamma(t)
 \sigma_-.
\end{eqnarray}
To obtain the operator $\Omega(t)$ introduced in
Eq.~(\ref{OMEGA-DEF}) we first have to determine the quantity
$a=a(t)$ which is defined to be the largest eigenvalue of
$(C-D)^{\dagger}(C-D)$. This operator is equal to $0$ for
$\gamma(t)\geq 0$, and equal to $2|\gamma(t)|\sigma_+\sigma_-$ for
$\gamma(t)<0$. Thus we find
\begin{equation} \label{DEF-CD}
 (C-D)^{\dagger}(C-D) = a\sigma_+\sigma_-
\end{equation}
and
\begin{equation} \label{DEF-a}
 a(t) = |\gamma(t)|-\gamma(t).
\end{equation}
Hence, Eq.~(\ref{OMEGA-DEF}) takes the form
\begin{eqnarray}
 \Omega^{\dagger}\Omega &=& aI-(C-D)^{\dagger}(C-D)
 = a(I-\sigma_+\sigma_-) \nonumber \\
 &=& a\sigma_-\sigma_+,
\end{eqnarray}
which leads to an obvious solution:
\begin{equation} \label{DEF-omega}
 \Omega(t) = \sqrt{a(t)}\sigma_+.
\end{equation}
The Lindblad operators $J_i$ defined in
Eqs.~(\ref{DEF-J1})-(\ref{DEF-J4}) are therefore given explicitly
by:
\begin{eqnarray}
 J_1 &=& \sqrt{\frac{|\gamma|}{2}} \sigma_- \otimes
 \big[ |1\rangle\langle 1| + {\textrm{sign}}\gamma |2\rangle\langle 2| \big],
 \label{J1} \\
  J_2 &=& \sqrt{\frac{|\gamma|}{2}} \sigma_- \otimes
 \big[ {\textrm{sign}}\gamma |1\rangle\langle 1| + |2\rangle\langle 2| \big],
 \label{J2} \\
 J_3 &=& \sqrt{a}\sigma_+ \otimes |3\rangle\langle 1|,
 \label{J3} \\
 J_4 &=& \sqrt{a}\sigma_+ \otimes |3\rangle\langle 2|.
 \label{J4}
\end{eqnarray}

\subsection{Physical interpretation}\label{INTER}
Considering times $t$ for which $\gamma(t)\geq0$, we have $a=0$
[see Eqs.~(\ref{DEF-a})] and, hence, $J_3=J_4=0$. This means that
for $\gamma(t)\geq0$ the decay channels described by $J_3$ and
$J_4$ are closed: The process only involves the jumps of the state
vector given by the operators $J_1$ and $J_2$. We infer from
Eqs.~(\ref{J1}) and (\ref{J2}) that $J_1$ and $J_2$ induce
downward transitions $|e\rangle\rightarrow|g\rangle$ of the two
state system (action of $\sigma_-$). These transitions result from
the projection of the system's state vector into the ground state
$|g\rangle$ conditioned on the detection of a quantum in reservoir
$R_1$ or $R_2$.

For $\gamma(t)<0$ the operators $J_1$ and $J_2$ again induce
downward transitions of the two state system and, at the same
time, introduce a relative phase factor of ${\textrm{sign}}\gamma
= -1$ between the states $|1\rangle$ and $|2\rangle$ of the
auxiliary three state system. Moreover, in the case $\gamma(t)<0$
the decay channels $J_3$ and $J_4$ are open: This enables the
additional jumps of the state vector described by $J_3$ and $J_4$.
Equations (\ref{J3}) and (\ref{J4}) show that $J_3$ and $J_4$ lead
to upward transitions $|g\rangle\rightarrow|e\rangle$ of the two
state system (action of $\sigma_+$) with simultaneous transitions
between the auxiliary states of the form
$|1\rangle\rightarrow|3\rangle$ or
$|2\rangle\rightarrow|3\rangle$. A re-population of the excited
state $|e\rangle$ is thus possible through jumps into the
auxiliary state $|3\rangle$, corresponding to the detection of a
quantum in reservoir $R_3$ or $R_4$.

A detailed analysis of the process can be given in terms of the
statistics of the quantum jumps. To this end, we note that the
waiting time distribution for the SSE (\ref{SSE}) is given by
\begin{equation} \label{WDF}
 F(t_1,t_0) = 1 -
 \left|\left| T \exp\left[ -i\int_{t_0}^{t_1} ds
 \hat{H}(s)\right] |\Phi(t_0)\rangle \right|\right|^2,
\end{equation}
where $\hat{H}$ is defined by Eq.~(\ref{H-HAT}). $F(t_1,t_0)$ is
the probability that a jump takes place in the time interval
$(t_0,t_1)$, given that the previous jump occurred at time $t_0$
and yielded the state $|\Phi(t_0)\rangle$. Note that this is a
true cumulative probability distribution, i.e. $F(t_1,t_0)$
increases monotonically with $t_1$ and satisfies $F(t_0,t_0)=0$.
For our model we have [see Eqs.~(\ref{H-HAT}), (\ref{DEF-H}),
(\ref{SUMI-JI}) and (\ref{OMEGA-EQ})]:
\begin{eqnarray} \label{H-HAT-JC}
 \hat{H}(t) &=& H_S(t) \otimes I_3 \\
 &~& -\frac{i}{2}\left(a(t)I+\gamma(t)\sigma_+\sigma_-\right)\otimes
 \left(|1\rangle\langle 1| + |2\rangle\langle 2| \right).
 \nonumber
\end{eqnarray}

Let us analyze the process starting from the initial state
\begin{equation} \label{INIT-JC}
 |\Phi(0)\rangle=|e\rangle\otimes|\chi\rangle,
\end{equation}
and investigate the occupation probability $p_g(t)$ of the ground
state. From the master equation it is clear that this quantity is
given by the simple expression:
\begin{equation} \label{P-E-EXACT}
 p_g(t) = 1-\exp\left[-\int_0^t ds \gamma(s)\right].
\end{equation}
If $\gamma(s)$ takes on positive and negative values, this is a
non-monotonic function of time. Our aim is to illustrate how the
stochastic dynamics reproduces this behavior and to explain the
physical picture provided by the unraveling.

In the stochastic representation we have the formula [see
Eq.~(\ref{RHO-STOCH})]:
\begin{equation} \label{P-E}
 p_g(t) =
 \frac{E[\langle g|\psi_1(t)\rangle\langle\psi_2(t)|g\rangle]}
 {E[\langle\psi_2(t)|\psi_1(t)\rangle]}.
\end{equation}
We denote the moment of the first jump by $t_1$, the moment of the
second jump by $t_2$. It follows from Eq.~(\ref{WDF}) that the
waiting time distribution for the first jump is given by
\begin{equation}
 F(t_1,0) = 1 - \exp\left[-\int_0^{t_1} ds |\gamma(s)|\right],
\end{equation}
and that the waiting time distribution for the second jump, given
that the first jump took place at time $t_1$, becomes
\begin{equation}
 F(t_2,t_1) = 1 - \exp\left[-\int_{t_1}^{t_2} ds a(s)\right].
\end{equation}

We further denote the total number of jumps in the time interval
$(0,t)$ by $N(t)$. Since the process starts from the state
(\ref{INIT-JC}) the first jump is given by the application of
$J_1$ or $J_2$ which project the state vector into the ground
state. Therefore, prior to the first jump we have $2\langle g
|\psi_1(t)\rangle\langle\psi_2(t)|g\rangle=0$, immediately
afterwards we get $2\langle g
|\psi_1(t)\rangle\langle\psi_2(t)|g\rangle={\mathrm{sign}}\gamma(t_1)$.
During the time intervals in which $\gamma(t)<0$ a second jump
described by $J_3$ or $J_4$ is possible by which the state vector
leaves the manifold ${\mathcal{H}}_1\oplus{\mathcal{H}}_2$ and
lands in ${\mathcal{H}}_3$. Once the state vector is in
${\mathcal{H}}_3$, no further jumps are possible. Of course, we
get again $2\langle g
|\psi_1(t)\rangle\langle\psi_2(t)|g\rangle=0$ after the second
jump. Summarizing we have three possible alternatives:
\begin{eqnarray}
 N(t) = 0 &\Rightarrow&
 2\langle g |\psi_1(t)\rangle\langle\psi_2(t)|g\rangle = 0, \\
 N(t) = 1 &\Rightarrow&
 2\langle g |\psi_1(t)\rangle\langle\psi_2(t)|g\rangle
 = {\mathrm{sign}}\gamma(t_1), \label{XSIGMA} \\
 N(t) = 2 &\Rightarrow&
 2\langle g |\psi_1(t)\rangle\langle\psi_2(t)|g\rangle = 0.
\end{eqnarray}
From these relations we find the expectation value
\begin{eqnarray} \label{X-T}
 \lefteqn{
 E[2\langle g |\psi_1(t)\rangle\langle\psi_2(t)|g\rangle]
 } \nonumber \\
 &=& \int_0^t dt_1 \dot{F}(t_1,0)[1-F(t,t_1)]
 {\mathrm{sign}}\gamma(t_1) \\
 &=& \exp\left[-\int_0^t dt_1 a(t_1)\right]
 -\exp\left[-\int_0^t dt_1 |\gamma(t_1)| \right]. \nonumber
\end{eqnarray}
Here, the quantity $dt_1\dot{F}(t_1,0)$ is the probability that
the first jump occurs in $dt_1$, while $1-F(t,t_1)$ is the
probability that no further jumps take place within $(t_1,t)$.

A similar analysis can be performed to obtain the expectation
value of the quantity $2\langle\psi_2(t)|\psi_1(t)\rangle$. One
finds:
\begin{eqnarray}
 N(t) = 0 &\Rightarrow& 2\langle\psi_2(t)|\psi_1(t)\rangle
 = 1, \\
 N(t) = 1 &\Rightarrow& 2\langle\psi_2(t)|\psi_1(t)\rangle
 = {\mathrm{sign}}\gamma(t_1), \label{YSIGMA1} \\
 N(t) = 2 &\Rightarrow& 2\langle\psi_2(t)|\psi_1(t)\rangle
 = 0. \label{YSIGMA2}
\end{eqnarray}
Thus we get:
\begin{eqnarray} \label{Y-T}
 \lefteqn{
 E[2\langle\psi_2(t)|\psi_1(t)\rangle] } \nonumber \\
 &=& 1-F(t,0) + \int_0^t dt_1 \dot{F}(t_1,0)[1-F(t,t_1)]
 {\mathrm{sign}}\gamma(t_1) \nonumber \\
 &=& \exp\left[-\int_0^t dt_1 a(t_1)\right].
\end{eqnarray}
The term $1-F(t,0)$ represents the no-jump probability, i.~e., the
contribution from the event $N(t)=0$. The result (\ref{Y-T}) could
have been obtained also directly from Eq.~(\ref{TRACE}). Using
finally Eqs.~(\ref{X-T}) and (\ref{Y-T}) in Eq.~(\ref{P-E}) we
find, of course, the correct expression (\ref{P-E-EXACT}) for the
excited state probability.

\begin{figure}[htb]
\includegraphics[width=\linewidth]{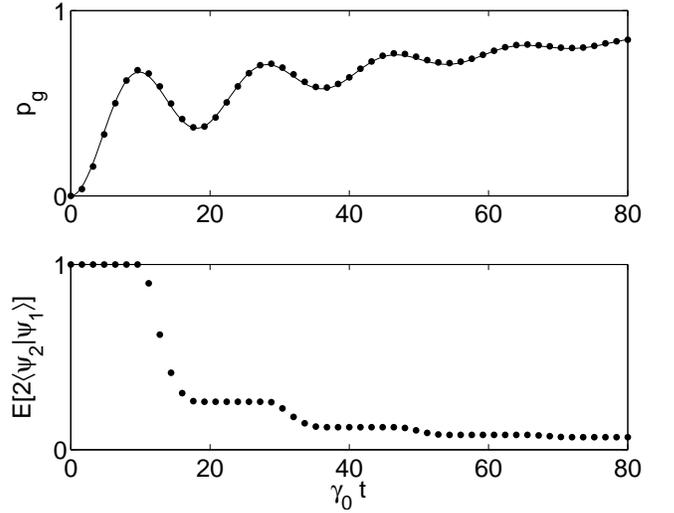}
\caption{Simulation of the SSE (\ref{SSE}) for the non-Markovian
decay of a two-state system. Top: ground state probability $p_g$
obtained from a sample of $10^5$ quantum trajectories (dots), and
analytical solution (continuous line). Bottom: expectation value
$E[2\langle\psi_2|\psi_1\rangle]$. Parameters:
$\gamma_0/\lambda=25$, $\Delta/\gamma_0=0.2$. \label{fig3}}
\end{figure}

We illustrate the above analysis by means of an example. Figure
\ref{fig3} shows the results of a Monte Carlo simulation of the
SSE (\ref{SSE}) for the case of a Lorentzian spectral density
which is detuned from the transition frequency of the two state
system by an amount $\Delta$ (damped Jaynes-Cummings model). This
leads to a bath correlation function of the form
\begin{equation} \label{CORR-JC}
 \langle 0|Q(t)Q^{\dagger}(t_1)|0\rangle =
 \frac{\gamma_0\lambda}{2}
 e^{i\Delta (t-t_1) - \lambda |t-t_1|},
\end{equation}
where $\gamma_0^{-1}$ is the Markovian relaxation time and
$\lambda^{-1}$ is the correlation time of the bath. The simulation
was carried out in a nonperturbative regime: While the Born-Markov
approximation requires that $\gamma_0/\lambda \ll 1$, the
simulation uses $\gamma_0/\lambda = 25$. In this regime
$\gamma(t)$ becomes negative for certain time intervals. These
intervals can be seen in the figure as those intervals over which
$p_g$ and $E[2\langle\psi_2|\psi_1\rangle]$ decrease monotonically
with time. This is a signature for the fact that transitions
$|1\rangle\rightarrow|3\rangle$ or $|2\rangle\rightarrow|3\rangle$
become possible through the channels $J_3$ or $J_4$. These
channels are closed in the time intervals over which
$E[2\langle\psi_2|\psi_1\rangle]$ stays constant.

The decrease of the ground state probability can be interpreted as
due to virtual processes in which a quantum is emitted into the
bath and re-absorbed at a later time. This is a clear
non-Markovian feature of the dynamics. In the stochastic
unraveling this decrease results from the contributions of those
quantum trajectories which involve at least one jump and for which
the first jump at time $t_1$ occurred during a phase in which
$\gamma(t_1)<0$. The first jump then yields a negative
contribution to the expectation values of $2\langle
g|\psi_1\rangle\langle\psi_2|g\rangle$ and of
$2\langle\psi_2|\psi_1\rangle$ as a result of the relative phase
factor ${\mathrm{sign}}\gamma(t_1)=-1$ between the states
$|1\rangle$ and $|2\rangle$ introduced by $J_1$ and $J_2$ [see
Eqs.~(\ref{XSIGMA}) and (\ref{YSIGMA1})]. Moreover, in a possible
second jump the state vector leaves the manifold
${\mathcal{H}}_1\oplus{\mathcal{H}}_2$ to end up in a state
proportional to $|e\rangle\otimes|3\rangle$, which gives
$2\langle\psi_2|\psi_1\rangle=0$ [see Eq.~(\ref{YSIGMA2})].

The decrease of $p_g$ is therefore due to quantum trajectories for
which a second jump is possible which leads to a re-excitation
$|g\rangle\rightarrow|e\rangle$. Thus we see that the virtual
emission and re-absorption processes appear in the stochastic
unraveling in the extended state space as certain real processes,
namely as jumps with $J_1$ or $J_2$ involving a negative phase
factor (detection of a quantum in reservoir $R_1$ or $R_2$), and
as jumps into the auxiliary state $|3\rangle$ (detection of a
quantum in reservoir $R_3$ or $R_4$). This shows how the quantum
memory effect of virtual emission and re-absorption processes is
encoded and completely stored in the continuous measurement
record.

\section{Discussion and conclusions}\label{CONCLU}
In this paper we have developed a general method for the
derivation of stochastic unravelings for non-Markovian quantum
processes given by time-local master equations of the form
(\ref{GEN-NON-MARKOV}). The key point of the construction is the
fact that such master equations always allow a Markovian embedding
in an extended state space with a rather simple structure, namely
in the triple Hilbert space ${\mathcal{H}}\otimes{\mathbb C}^3$.
Within this embedding the density matrix $\rho(t)$ of the original
open system is expressed through a certain set of coherences of
the full density matrix $W(t)$ on the extended state space.

The transition to the extended state space can be viewed
physically as the addition of a further degree of freedom which is
realized by a three level system. This enables one to represent
the given non-Markovian dynamics by means of a suitable
interaction with a Markovian environment consisting of the
reservoirs $R_i$ introduced in Sec.~\ref{CMT-MARKOV}. Although the
generator of the given non-Markovian master equation needs not be
in Lindblad form, the corresponding dynamics in the extended state
space is therefore governed by a time dependent Lindblad generator
of the form of Eq.~(\ref{LINDBLAD-EQ}). The lifting to an
appropriate extended state space thus allows the derivation of
stochastic Schr\"odinger equations for non-Markovian dynamics
through a consistent application of the standard theory of quantum
measurement. The SSEs obtained in this way generate genuine
quantum trajectories with the physical interpretation of
continuous measurements.

The construction of Sec.~\ref{EMBEDDING} provides a fairly general
method for the Markovian embedding of a given non-Markovian
dynamics: Apart from the existence of the master equation and from
the boundedness of the operators $C_{\alpha}(t)$ and
$D_{\alpha}(t)$, no assumption was made regarding the interaction
Hamiltonian, the spectral density, the reservoir state, its
temperature, etc. The Markovian embedding could therefore be
useful in itself since it enables one to employ well established
and developed concepts from the theory of completely positive maps
and Lindblad generators in the study of non-Markovian master
equations.

Formulating a non-Markovian unraveling we made use of piecewise
deterministic jump processes. In an electromagnetic environment
this corresponds, for example, to direct photodetection. It should
be clear, however, that our derivation allows any unraveling in
the extended state space. Alternatively one can use diffusion-type
SSEs, which in a continuous measurement interpretation correspond
to other detection schemes like homodyne or heterodyne
photodetection \cite{WISEMAN1,WISEMAN2}.

Various stochastic unravelings for non-Markovian dynamics have
been suggested in the literature, involving both jump processes
\cite{IMAMOGLU,JACK} as well as SSEs with colored noise
\cite{DGS1,DGS2,GAMBETTA1,GAMBETTA2}. The technique developed by
Imamoglu \cite{IMAMOGLU} is related to the method of pseudo modes
\cite{GARRAWAY1,GARRAWAY2}. It employs an approximate Markovian
embedding of a given non-Markovian dynamics. This embedding is
based on the assumption that the reservoir can be represented by
means of an effective set of fictitious damped harmonic oscillator
modes. The Markovian embedding of the present paper is realized in
an entirely different way by the introduction of the triple
Hilbert space, and avoids the expansion into pseudo modes. It
should also be noted that in the present method the density matrix
$\rho(t)$ is \textit{not} given by the partial trace of the
density matrix $W(t)$ in the extended state space.

A further interesting method has been formulated by Di\'{o}si,
Gisin and Strunz \cite{DGS1,DGS2}. These authors employ a nonlocal
stochastic integro-differential equation for the state vector. As
demonstrated by Gambetta and Wiseman \cite{GAMBETTA1} it seems,
however, that the nonlocal SSE does not admit a continuous
measurement interpretation within the framework of standard
quantum measurement theory (see also \cite{GAMBETTA3} in this
context). This means that measurements carried out at different
times on the environment will influence the dynamics in a way
which is incompatible with the stochastic process. The SSE does
therefore not generate genuine quantum trajectories in the sense
it does for Markovian dynamics.

A number of unravelings of non-Markovian dynamics has been
proposed \cite{BKP,STOCK} which are based on the idea of
propagating a pair $|\psi_1(t)\rangle$, $|\psi_2(t)\rangle$ of
stochastic state vectors and of representing the reduced density
matrix with the help of the expectation value
$\rho(t)=E[|\psi_1(t)\rangle\langle\psi_2(t)|]$. It is even
possible to design an exact stochastic unraveling
\cite{PDP-EPJD,PDP-PRA} which neither requires the existence of a
master equation nor a factorizing initial state. This method makes
use of a pair of independently evolving product states in the
state space of the total system. Related stochastic wave function
methods have also been formulated for the description of bosonic
and fermionic many-body systems \cite{CARUSO,CHOMAZ}, and for the
simulation of quantum gases \cite{STEEL,CORNEY} by use of the
positive P-representation \cite{DRUMMOND,GILCHRIST}. A measurement
interpretation of these stochastic methods is however not
available.

A pair $|\psi_1(t)\rangle$, $|\psi_2(t)\rangle$ of state vectors
can be considered as an element of the double Hilbert space
${\mathcal{H}}\otimes{\mathbb
C}^2\cong{\mathcal{H}}_1\oplus{\mathcal{H}}_2$ which is the tensor
product of ${\mathcal{H}}$ and the state space of a two state
system. In Ref.~\cite{BKP} a stochastic unraveling in the double
Hilbert space has been constructed. Although this method has been
demonstrated to provide a useful numerical tool, a continuous
measurement interpretation seems again to be impossible. This is
connected to the facts that not only the master equation in
${\mathcal{H}}$ but also the master equation in the double Hilbert
space is generally not in Lindblad form and that the process does
not preserve the norm of the state vector.

We mention finally some restrictions of the present theory.
Similar to the formulation of the Lindblad theorem, we made use of
the assumption of boundedness of the operators $C_{\alpha}(t)$ and
$D_{\alpha}(t)$ in the non-Markovian master equation
(\ref{GEN-NON-MARKOV}). This assumption excludes the immediate
treatment of important cases, such as quantum Brownian motion
which involves the unbounded operators for position and momentum
of the particle. However, what is really needed in the proof is
that the inequality (\ref{INEQ}) is satisfied. Provided the
dynamics of the state vector is confined to an effective subspace
of ${\mathcal{H}}$ on which the right-hand side of inequality
(\ref{INEQ}) is bounded, we can still define a finite $a(t)$ and
construct the embedding. A further restriction of the theory is
that for certain models time-local master equation of the form
(\ref{GEN-NON-MARKOV}) may not exist for very strong couplings.
The latter can lead to singularities of the TCL generator and to a
breakdown of the TCL expansion (an example is discussed in
\cite{TheWork}). It is an important open problem whether a
continuous measurement unraveling can be developed for such cases.

\begin{acknowledgments}
The author would like to thank D. Burgarth and F.~Petruccione for
helpful comments and stimulating discussions.
\end{acknowledgments}

\end{document}